\documentclass[lettersize,journal]{IEEEtran}
\usepackage{amsmath,amsfonts}
\usepackage{array}
\usepackage[caption=false,font=normalsize,labelfont=sf,textfont=sf]{subfig}
\usepackage{textcomp}
\usepackage{stfloats}
\usepackage{url}
\usepackage{verbatim}
\usepackage{graphicx}
\graphicspath{{Figures/Figures_PDF}}
\usepackage[flushleft]{threeparttable}
\def\BibTeX{{\rm B\kern-.05em{\sc i\kern-.025em b}\kern-.08em
    T\kern-.1667em\lower.7ex\hbox{E}\kern-.125emX}}
\usepackage{balance}
\begin{document}
\title{Fine-Resolution Silicon Photonic Wavelength-Selective Switch Using Hybrid Multimode Racetrack Resonators}
\author{Lucas M. Cohen, Saleha Fatema, Vivek V. Wankhade, Navin B. Lingaraju, Bohan Zhang, Deniz Onural, Milo{\v s} Popovi{\'c}, and Andrew M. Weiner
\thanks{L. M. Cohen, S. Fatema, V. V. Wankhade, and A. M. Weiner are with the Purdue Quantum Science and Engineering Institute and the Elmore Family School of Electrical and Computer Engineering, Purdue University, West Lafayette, IN 47907, USA. 

Navin B. Lingaraju is with SRI International, Arlington, VA 22209, USA. His present address is the John Hopkins University Applied Physics Laboratory, 11100 Johns Hopkins Rd, Laurel, MD 20723, USA. 

Bohan Zhang, Deniz Onural, and Milo{\v s} Popovi{\'c} are with the Department of Electrical and Computer Engineering,, Boston University, Boston, MA 02215, USA.}}

\maketitle

\begin{abstract}
In this work, we describe a procedure for synthesizing racetrack resonators with large quality factors and apply it to realize a multi-channel wavelength-selective switch (WSS) on a silicon photonic chip. We first determine the contribution of each component primitive to propagation loss in a racetrack resonator and use this data to develop a model for the frequency response of arbitrary order, coupled-racetrack channel dropping filters. We design second-order racetrack filters based on this model and cascade multiple such filters to form a 1$\times$7 WSS. We find good agreement between our model and device performance with second-order racetrack that have  $\approx$~1~dB of drop-port loss, $\approx$~2~GHz FWHM linewidth, and low optical crosstalk due to the quick filter roll-off of $\approx$~5.3~dB$/$GHz. Using a control algorithm, we show three-channel operation of our WSS with a channel spacing of only 10~GHz. Owing to the high quality factor and quick roll-off of our filter design, adjacent channel crosstalk is measured to be $<$~$-$25 dB for channels spaced on a 10~GHz grid. As a further demonstration, we use five of seven WSS channels to perform a demultiplexing operation on both an 8~GHz and a 10~GHz grid. These results suggest that a low-loss WSS with fine channel resolution can be realized in a scalable manner using the silicon photonics platform.
\end{abstract}


\section{Introduction}
\label{sec1}

\begin{figure*}[t]
\centering
\includegraphics[scale=0.54]{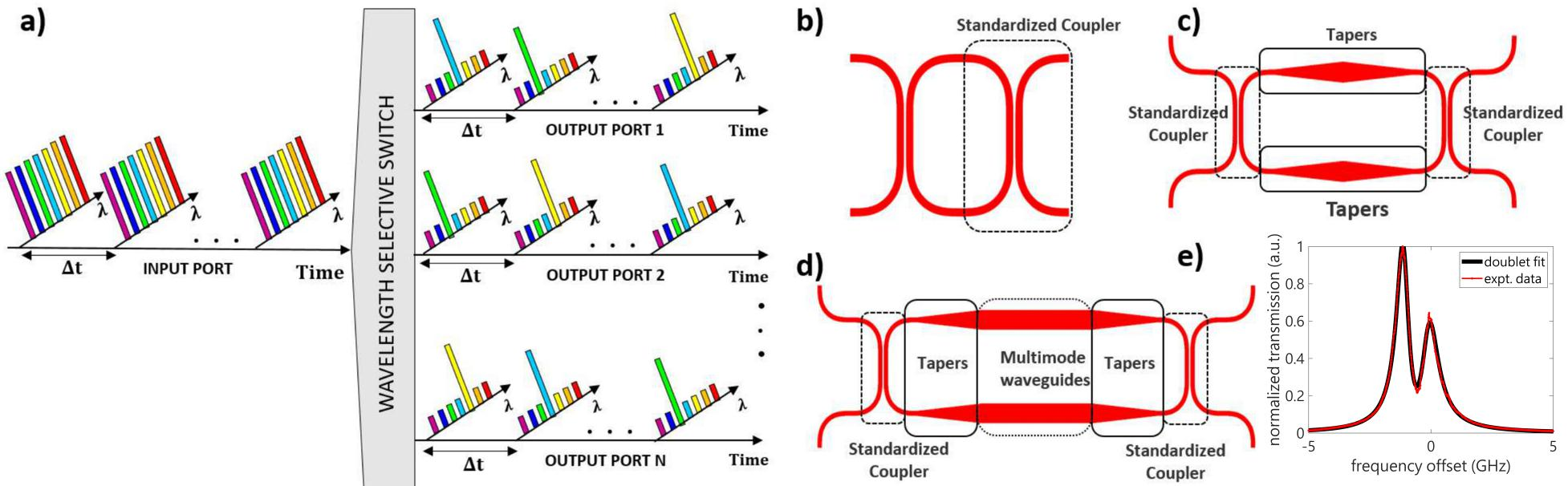} 
\caption{ 
(a) Conceptual diagram of a 3-channel WSS.
(b) Weakly coupled resonator to characterize standardized couplers.
(c) Weakly coupled resonator to characterize adiabatic tapers.
(d) Full hybrid multimode racetrack filter comprised of single-mode standardized couplers, adiabatic single to multimode waveguide tapers, and multimode waveguides.
(e) Example spectral response from a weakly coupled structure with the geometry of Fig. \ref{fig1}(b) with a doublet resonance mode and fit.
}
\label{fig1}
\end{figure*}

\IEEEPARstart{A}{s} internet traffic volume and the demand for data increases, existing optical transport architectures will require hardware upgrades for the improvement of transmission capacity and capabilities of optical networks. Reconfigurable optical add-drop multiplexers (ROADMs) are a critical building block of optical networks, enabling flexibility in wavelength routing and assignment between users on an optical network. ROADMs are comprised principally of a wavelength-selective switch (WSS, shown schematically in Fig. \ref{fig1}(a)), and WSSs that are actively deployed today are most commonly based on diffraction grating based spectral dispersers and liquid crystal on silicon (LCoS) technology \cite{gringeri}.  Although well suited to optical network requirements, such WSSs require assembly of bulk or micro-optic components,  are  typically limited to spectral resolutions at ca. 10 GHz and above, and induce excessive optical loss especially when modified to perform filtering operations at unusually fine spectral resolutions in the few GHz range \cite{ma}.

Due to the inherent scalability, low-cost, and robust component catalog of silicon photonics technology (SiP), there has been an increasing research effort towards realizing a SiP WSS \cite{siew}. A promising approach for a SiP WSS is to use microresonators as a filtering element due to their resonant selectivity, compact footprint, and simple operation \cite{feng, bogaerts}. An $M$ input spatial channel, $L$ output wavelength channel $M$$\times$$L$ WSS \cite{zhang}, a flexible-grid WSS \cite{chen}, and even an $M$ input spatial channel, $N$ output spatial channel, $L$ wavelength channel/port $M$$\times$$N$$\times$$L$ WSS \cite{khope} have been demonstrated using microresonators with SiP technology. However, current demonstrations have used microresonators with relatively low quality factors, rendering them incapable of meeting the fine-resolution requirements that could be asked of ROADMs in the future \cite{perrin}. For example, over the past decade there has been a push towards utilizing optical superchannels to improve spectral efficiency and total throughput in an optical network \cite{winzer, marom}. There are scenarios in which subchannel add/drop capabilities within a superchannel are desirable. Performing these functions optically has a number of potential benefits to the network, but it requires high-selectivity filtering (resolution at the single GHz level) to e.g. add guard bands between adjacent subchannels as well as to assist in picking out the subchannel. Although there have been impressive demonstrations of such hyperfine resolution filtering \cite{klonidis,rudnick,xu,xiao,Metcalf}, such demonstrations rely on substantially more complex optical setups or exotic components and are subject to increased insertion loss, as mentioned above.

Selective wavelength filtering (sub-GHz linewidths) is also important in applications like quantum information science and microwave photonics. In particular, many view modular architectures as essential to the scale-up of quantum computing systems, where communications and entanglement between individual computational modules is mediated by photons \cite{duan}.
These photons, which must interface with matter-based qubits, will have linewidths on the order of 10s to 100s of MHz. Consequently, low-loss and narrowband filters can help selectively manipulate or route these modes over local area or larger networks \cite{quantinuum}.
Microwave photonics, the science of processing radiofrequency signals in the optical domain \cite{marpaung}, similarly requires the ability to control and isolate narrowband optical signals with low loss in order to realize systems with low noise figures. Such systems can be harnessed for applications like abritrary waveform generation and shaping of low-repetition rate combs, among others \cite{wang}.

In this paper, we propose and experimentally demonstrate a multi-channel WSS using racetrack resonators with a hybrid geometry that includes wide waveguide segments to make possible optical filtering at a fine-resolution. To this end, we first develop a methodology for the robust modeling of arbitrary order coupled-racetrack devices using experimental data. The developed model shows good agreement with experimental results and enables first-time-right designs of narrowband filters. Next, we use the model to design a 1$\times$7 WSS with a second-order filter response with $\approx$~1~dB of drop-port loss, 2~GHz FWHM channel linewidth, and 5.3~dB$/$GHz of roll-off, and we demonstrate WSS operation using 3 filter channels. A subset of results from this manuscript was presented at the IEEE Photonics Conference \cite{IPC_synthesis} and Conference on Lasers and Electro-Optics \cite{Cleo_wss}. Here, we significantly extend the design methodology and results from previous proceedings. Our work takes a step towards showing the SiP platform is capable of meeting the fine-resolution filtering requirements of future-generation optical networks.

\section{Filter Model}
\label{sec2}

Our racetrack filters comprise three elements: a single-mode coupling region, adiabatic tapers from 0.5 $\mu m$ wide to 2 $\mu m$ wide waveguides, and 2 $\mu m$ wide multimode waveguides. Adiabatic tapers facilitate the propagation of the fundamental mode from a single-mode waveguide to multimode without excitation or loss to higher-order modes. The multimode waveguide, when operated in the fundamental mode, significantly reduces the dominant loss mechanism for SiP waveguides of field-sidewall overlap, thus providing low-loss \cite{onural2}. For sufficiently long multimode waveguide sections, the average round-trip loss through the racetrack is dominated by this element, thereby enabling one to flexibly tune the resonator's quality factor with the multimode waveguide length \cite{bogaerts}.

To characterize the loss contributions of each element of our racetrack resonators, we design structures on a full-stack active multi-project wafer (MPW) run \cite{IPC_synthesis}.
These structures follow the weakly-coupled cavity method \cite{popovic} in which a resonator is sufficiently weakly coupled to bus waveguides such that the linewidth of its frequency response is dominated by the intrinsic loss of the resonator
as opposed to the external losses from the coupling to the bus waveguides. A careful balance must be achieved so as to be sufficiently weakly coupled with the resonator yet  coupled enough to have a strong signal at the drop port. In this way, the loss contributions from an arbitrary component can be measured without taking up an excessive footprint on the chip.

The three classes of test structures are shown in Fig. \ref{fig1}(b-d). Each consists of a resonator comprised of a pair of what we term \textit{standardized couplers} that sandwich either two pairs of tapers or two pairs of tapers and a pair of multimode waveguides that we want to characterize. The standardized coupler is a single-mode region formed by two 90$^{\circ}$ graduated radii bends with a 5~$\mu m$ long straight waveguide in between. This straight coupling section provides a long interaction region to realize sufficiently high bus-racetrack coupling even for large bus-racetrack gaps, which are better tolerated by the fabrication process and can also lead to reduced coupling losses. The 90$^{\circ}$ bends are hybrid structures comprising two Euler curves and a circular arc of constant radius that matches the minimum radius of the Euler curves \cite{bahadori}. In this way, the Euler curve reduces mode mismatch loss from the interface with a straight waveguide, while the circular section minimizes the footprint of the composite curve. These standardized couplers simplify the formation of coupled-resonator designs since the physical coupling structure is identical for both bus-resonator and resonator-resonator couplings.

These test structures were fabricated through an active MPW run through AIM Photonics on 220 nm-thick silicon-on-insulator (SOI) wafers \cite{fahrenkopf}. 
There are devices for three variations of the standardized coupler - variants with minimum radii ($R_{\rm min}$) of 3 $\mu m$, 4 $\mu m$, and 5 $\mu m$. Our test devices covered a range of linear taper lengths from 30~$\mu m$ to 200~$\mu m$. Besides the linear tapers, we also included a set of full racetracks (Fig \ref{fig1}(d)) with adiabatic tapers where the tapered shape synthesized is based on a Fourier modal method so that the taper length is minimized for a given loss \cite{onural, song}. Also, we have two devices to measure 2~$\mu m$ wide waveguide loss with different 2~$\mu m$ waveguide lengths of 700~$\mu m$ and 1200~$\mu m$. These devices contain a 100~$\mu$m long linear taper to expand the waveguide to a 2~$\mu$m width. Finally, we have one test device with a weakly coupled circular microring resonator of radius $R = $~20~$\mu m$ with 0.5~$\mu m$ wide waveguides to gather a baseline single-mode waveguide loss for reference. All the test devices are designed with identical input and drop bus coupling gaps, and we vary this gap over a broad range such that we can successfully characterize a weakly coupled device.

\begin{table*}[t]
\caption{\bf Extracted Loss for Racetrack Resonators and Individual Subcomponents}
\begin{center}
\begin{tabular}{ccccccc}
\hline
Device & Structure of Interest  & Resonator &  Resonator &  Structure & Structure \\
 & (Units in $\mu$m) & Intrinsic Q &  Average Loss (dB/cm) & Loss (dB) &  Average Loss (dB/cm) \\
\hline
Weakly Coupled Microring        & Ring Resonator, ($R = $ 20)               & $4.37\times10^5$   & 1.8  & 0.0110 & 1.8 \\
Figure~\ref{fig1}(b)            & Standardized Coupler $R_{\rm min}$ $=$ 3  & $1.43\times10^5$   & 5.43 & 0.0099 & 5.43 \\
Figure~\ref{fig1}(b)            & Standardized Coupler $R_{\rm min}$ $=$ 4  & $2.33\times10^5$   & 3.32 & 0.0075 & 3.32 \\
Figure~\ref{fig1}(b)            & Standardized Coupler $R_{\rm min}$ $=$ 5  & $3.29\times10^5$   & 2.33 & 0.0064 & 2.33 \\
Figure~\ref{fig1}(c)            &   Linear Taper, Length $=$ 30             & $3.26\times10^5$   & 2.16 & 0.0062 & 2.09 \\
Figure~\ref{fig1}(c)            &   Linear Taper, Length $=$ 50             & $5.57\times10^5$   & 1.24 & 0.0047 & 0.95 \\
Figure~\ref{fig1}(c)            &   Linear Taper, Length $=$ 100            & $8.26\times10^5$   & 0.82 & 0.0062 & 0.62 \\
Figure~\ref{fig1}(c)            &   Linear Taper, Length $=$ 200            & $9.68\times10^5$   & 0.70 & 0.0117 & 0.59 \\
Figure~\ref{fig1}(d)            &   Fourier Taper$^{\ast}$, Length $=$ 11.36         & $1.64\times10^6$   & 0.40 & 0.0034 & 3.01 \\
Figure~\ref{fig1}(d)            &   Multimode Waveguide, Length $=$ 700     & $1.67\times10^6$   & 0.40 & 0.0170 & 0.24 \\
Figure~\ref{fig1}(d)            &   Multimode Waveguide, Length $=$ 1200    & $1.89\times10^6$   & 0.35 & 0.0300 & 0.25 \\

\hline
\end{tabular}
\end{center}
\vspace{-0.2cm} 
\begin{tablenotes}
            \item \hspace{0.5cm} $\ast$ Measured from device of Fig. \ref{fig1}(d) with 700~$\mu m$ long multimode waveguides.
\end{tablenotes}
  \label{tab}
\end{table*}

To couple light into the chips, we use a standard SMF-28 optical fiber array with an angled facet adjusted by an external polarization controller.
A representative spectrum with fitting from a single weakly-coupled device with the structure of Fig. \ref{fig1}(b) is shown in Fig \ref{fig1}(e). The average propagation loss of the racetrack resonator can be computed from the fitted quality factor and device parameters. The results are shown in Table 1, where we tabulate the intrinsic quality factor of racetrack resonators comprising different combinations of sub-components, as well as break out the contribution to loss from the sub-component or structure of interest \cite{IPC_synthesis}. In the table, the resonator intrinsic Q and average loss are computed directly from the frequency response of the device, while the contribution to loss from each sub-component is isolated after the contributions from previously characterized sub-components are removed.

We see that standardized couplers with $R_{\rm min} = $~5~$\mu m$ offered not only the lowest average loss, but also the lowest total structure loss, making them a good choice for multimode racetrack filters with high quality factors. For the linear tapers, a clear decrease in the taper's average loss is measured as the length of the taper increases, while the structure loss shows a more complicated trend. 
From the data in Table 1, we see that the Fourier tapers are measured to have the largest average loss but have the lowest insertion loss owing to their short length (11.36 $\mu m$). Finally, we computed the losses of 2~$\mu m$-wide multimode waveguides of different lengths. Because of their significant path length, they account for the dominant loss contribution in the racetrack yet their average loss of $\approx$~0.25~dB/cm is significantly lower than what we measure for a 0.5~$\mu m$ wide single mode waveguide of 1.8~dB/cm.

\begin{figure}[h]
\centering
\includegraphics[width=3.5in]{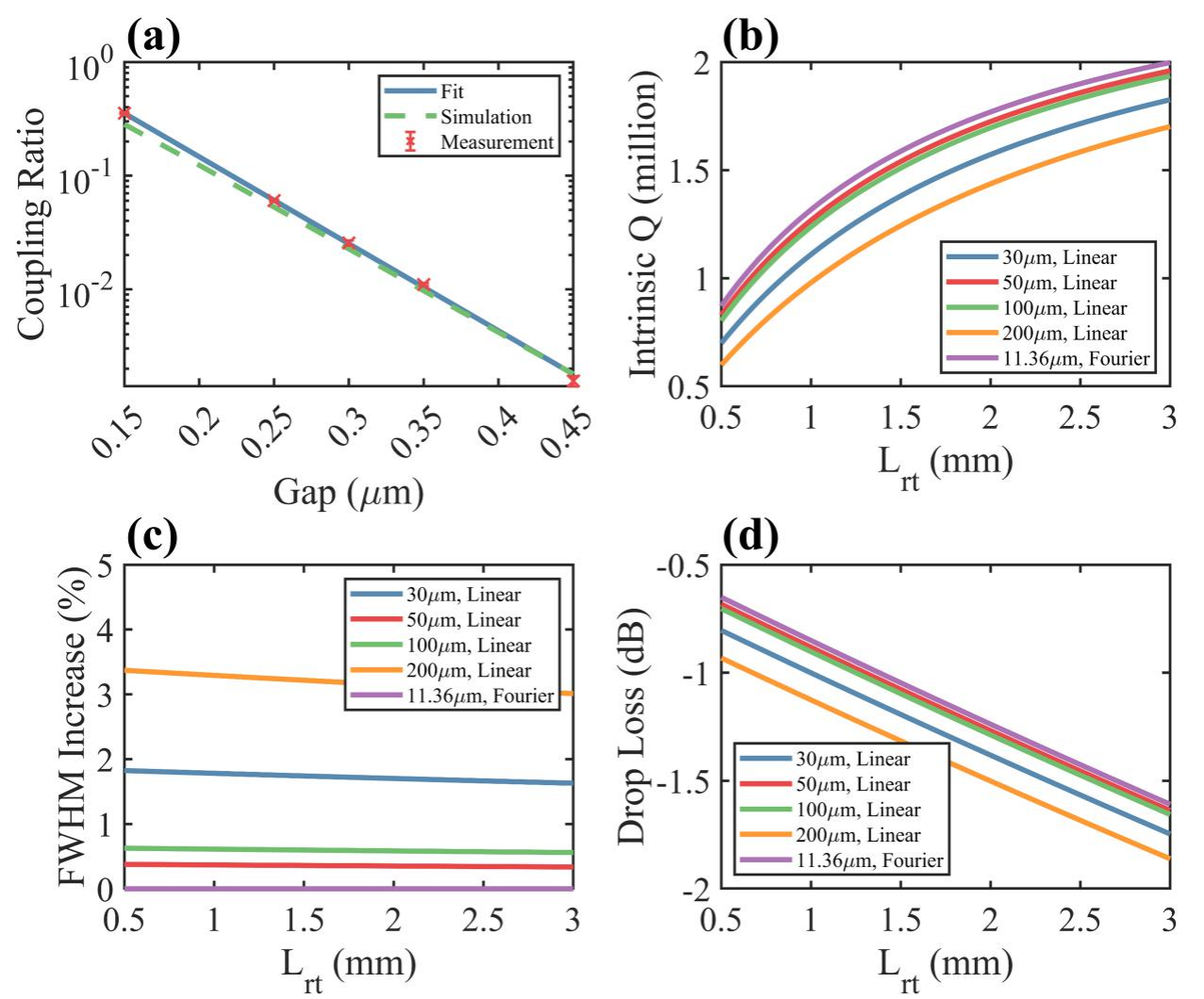}
\caption{ 
(a) Simulation and experimentally extracted cross coupling ratio versus gap for standardized couplers with $R_{\rm min}=$~5~$\mu m$.
Simulated (b) Intrinsic quality factor, (c) percent increase in 3dB linewidth relative to racetrack with Fourier tapers, (d) drop-port insertion loss versus round-trip racetrack resonator length for all measured taper variants, respectively, and assuming a constant power coupling of 5\%, respectively.
}
\label{fig2}
\end{figure}

In Fig. \ref{fig2}(a), we plot the power coupling ratio between standardized couplers of $R_{\rm min} =$~5~$\mu m$ as a function of the gap between them measured for the TE polarization. We are able to extract such information because of our variation in standardized coupler gaps from our test devices. Full 3-D finite-difference time domain (FDTD) simulation results show good agreement with what we extract from measurement, with a slight underestimate in simulation at smaller gaps ($<$~0.25~$\mu m$).

To quantitatively describe the frequency response of a racetrack comprised of a variation of these subcomponents, we developed a model using our experimentally extracted data. 
The model relies on well-known analytical formulas \cite{feng, bogaerts} and
is scalable to higher-order coupled racetrack structures thanks to the use of our standardized coupler geometry. 
An example of data from our model is shown in Fig. \ref{fig2} for a first-order racetrack designed with input and output bus waveguides having a coupling ratio of 5\%  (a gap near 0.25~$\mu m$ using our data from Fig. \ref{fig2}(a)).
We use the standardized coupler with $R_{\rm min} = $~5~$\mu m$ and plot a few useful quantities for a single resonance mode at $\lambda = $~1550nm for all our measured taper variants. The round-trip length of the racetrack is held fixed in the simulations by modifying the length of the multimode waveguide and therefore, for all taper variants to within a small error, the free spectral range (FSR) of the racetrack at a particular round-trip length is a constant. A group index of 3.5 is chosen in simulation and is near the measured value presented in Section \ref{sec3}.

From Fig. \ref{fig2}, we see that the Fourier tapers should be used to maximize the quality factor of the racetrack.
Interestingly, the intrinsic quality factor increases for linear tapers as we increase the length from 30~$\mu m$ to 50~$\mu m$, then it decreases slightly at 100~$\mu m$ and more rapidly for 200~$\mu m$. From this trend, we reason that the optimal linear taper length is somewhere between 50 and 100~$\mu m$ long. Longer than that and the taper expands too slowly for it to have a loss advantage.  
For all tapers, as the round-trip length increases, the resonance mode linewidth asymptotes towards a value around 500~MHz where light propagation in the racetrack becomes the dominant source of cavity loss. This value is based on our fixed coupling coefficient used in the simulation. However, the drop-port loss also increases quickly as the racetrack is now larger and has a greater total round-trip loss.

When making conclusions from our model towards designing useable racetrack devices, it is necessary to recognize the tradeoffs involving the resonator's FSR and its quality factor \cite{feng}. In particular, the intrinsic quality factor of a resonator is $Q_i = 2\pi n_{g} / \alpha_{avg}\lambda$ where $n_g$, $\alpha_{avg}$, and $\lambda$ are the group index, length-averaged round-trip loss, and resonance wavelength. The length-averaged round-trip loss can be expressed for our resonators as

\begin{equation}
\label{avg_loss}
\alpha_{avg} = \frac{\sum_{i=0}^{n}L_{i}\alpha_i}{\sum_{i=0}^{n}L_i}.
\end{equation}

where $\alpha_i$ and $L_i$ are the length-averaged loss and the length for the $i^{th}$ subcomponent of the resonator. In this form, it is clear that as the length of a particular component is increased, its contribution to the average loss of the resonator is as well. However, the FSR of the resonator, $\Delta \nu = c/ \sum_{i=0}^{n}L_i n_{g,i}$, is inversely proportional to its round-trip length. Hence, increasing the length of the lowest-loss component does indeed increase the quality factor of the resonator but at the cost of a reduced FSR, limiting the usable bandwidth of the device. Obviously, each parameter of the racetrack must be carefully chosen with the design application in mind.

\section{WSS Design and Operation}
\label{sec3}

\begin{figure*}[t]
\centering
\includegraphics[width=\textwidth]{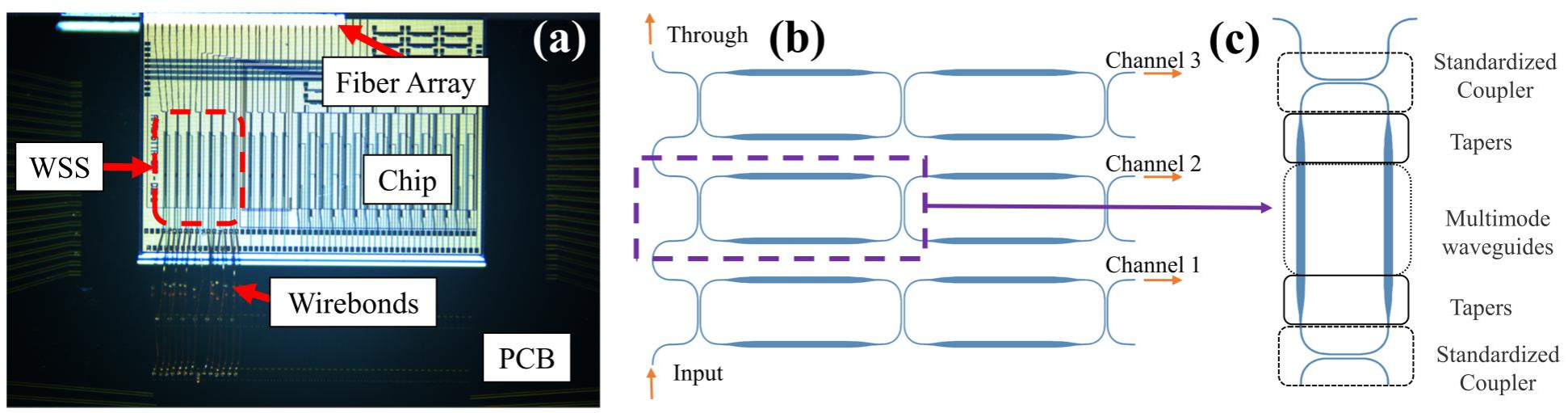}
\caption{ 
(a) Top-down microscope view of the silicon photonic WSS package.
(b) Layout of the on-chip WSS. 
(c) Zoom-in view of an individual racetrack of the coupled-resonator filter.
}
\label{fig3}
\end{figure*}

A 1$\times$N wavelength selective switch (schematically shown in Fig. \ref{fig1}(a) ) is a device which takes as input a broadband optical field and, in a re-programmable fashion, separates and routes distinct slices of that input spectrum to N distinct outputs. Our designed on-chip WSS structure is shown in \ref{fig3}(b). We use seven identical second-order racetrack filters coupled to a common input bus waveguide with unique drop-port I/O to form a 1$\times$7 WSS. Our racetrack filters are designed following the methodology described in Section \ref{sec2}, and we choose second-order structures for the increase in roll-off off of resonance. The bus-racetrack and racetrack-racetrack standardized coupler gaps are set to $\approx$~190~nm and 370~nm for power coupling coefficients of $\approx$~18 and 0.8~$\%$, respectively. In this way, we can pack filter channels more closely together than a single-order structure for the same level of inter-channel optical crosstalk. Each individual racetrack of the filter has the same round-trip length of $\approx$~1500~$\mu m$, uses Fourier tapers, and standardized couplers with $R_{\rm min} = $~5~$\mu m$. Standardized coupler gaps are set to achieve a flat passband frequency response with the smallest possible linewidth with the constraint that drop port loss does not exceed 1~dB.

The WSS chips were fabricated through an AIM Photonics MPW program at the SUNY Polytechnic Institute in a state-of-the-art 300mm facility. Photonics-grade 220~nm thick SOI wafers with a thick buried oxide are used with 193~nm immersion lithography to define silicon optical waveguiding structures. Embedded microheaters in the form of doped silicon slabs with various vias and metal interconnect layers for electrical contact are placed 1.25~$\mu$m from the optical waveguide in each racetrack resonator. In this way, an injected current in the slab can generate heat via the Joule heating effect to locally modify the refractive index and tune the racetrack resonance through silicon's thermo-optic effect. 

\begin{figure}[b!]
\centering
\includegraphics[width=8.5cm]{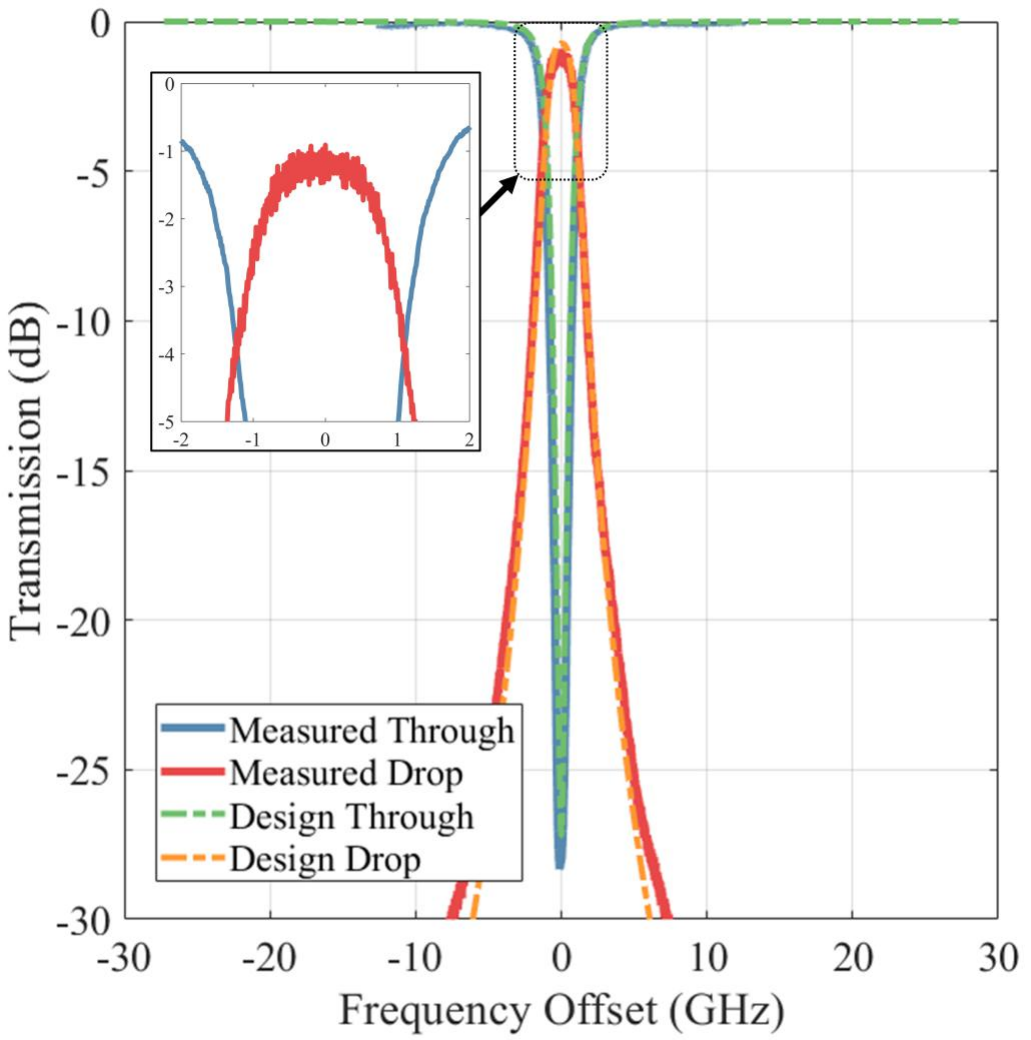}
\caption{ 
The measured optical frequency response for a single mode of the filter at $\lambda = $~1550 nm plotted with the designed filter response using our model developed in Section \ref{sec2}. The inset shows the flat-top passband.
}
\label{fig4}
\end{figure}

The fabricated chips were sent out for electrical wirebonding and packaging at the testing and packaging (TAP) facilities at Rochester Institute of Technology using a custom-developed electronic interposer and printed circuit board (PCB). 
To control each of the 7 WSS second-order racetrack filters, 21 wirebond connections were made.
The full footprint of the 1$\times$7 WSS is about 4 mm by 1.5 mm but can be significantly reduced by placing electrical and optical I/O more compactly. Excluding I/O, the 1x7 WSS occupies a footprint of 1.5 mm by 1.5 mm. The wirebonded package was placed on a custom temperature controlled chuck for compatibility with our optical probe station. A top-down view of the wirebonded chip package is shown in Fig. \ref{fig3}(a). The WSS was tested using a 16 channel SMF-28 optical fiber array with a flat facet fed by an Agilent 81632A tunable laser source with an external polarization controller. For thermal tuning, we use a 64-channel source measurement unit (SMU) with electronic cabling to interface with our package.

We first characterize the individual filter response of our WSS by sending laser light into the common through port and monitoring the output power at the common through port and drop port of channel 1. If we sweep the laser wavelength as we apply increasing electrical power to the thermo-optic heaters of a single racetrack of the filter, we can find the thermal state where the two resonators comprising the second-order filter are optically aligned. The final tuned state of the filter at a wavelength near 1550 nm is shown in Fig. \ref{fig4} along with a simulation using our model. We see good agreement between the two, further indicating the utility of the model. The passband shows a flat response with a $\approx$~2.1 GHz linewidth and 1 dB insertion loss. The FSR of the filter is about 54.5 GHz from which we extract an average group index of 3.54.

\begin{figure}[t]
\centering
\includegraphics[width=8.5cm]{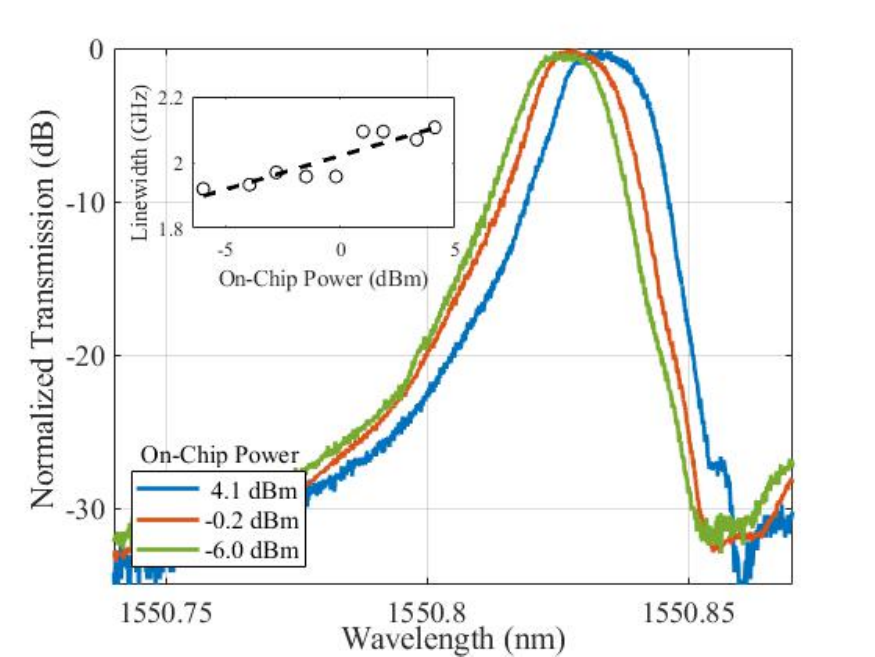}
\caption{ 
Frequency of a second-order racetrack filter for different on-chip input powers. The spectra are normalized with respect to the peak of each passband. The inset shows the change in the 3~dB linewidth of the filter as a function of optical power coupled onto the chip.
}
\label{fig4a}
\end{figure}

The performance of the second-order filter was also characterized over a range input optical powers to elucidate the impact of thermal effects \cite{carmon} and nonlinear absorption \cite{singer} on linewidth. Fig.~\ref{fig4a} shows the frequency response of this filter for on-chip input powers of -6.0~dBm, -0.2~dBm, and 4.1~dBm. The inset of Fig.~\ref{fig4a} shows the change in filter linewidth for nine different on-chip input power levels and we observe broadening that increases linearly with optical power in dBm over a range the spans input powers that span -6.0~dBm to 4.1~dBm. For powers above 4.1 dBm we began to observe bistability in the filter response.  

To use our system as a WSS, we need to tune each filter to its optimal response and then align them on a hypothetical frequency grid. To this end, we developed a Python algorithm with open-source libraries.
In the algorithm, a continuous-wave laser is input into the chip and repeatedly tuned between each frequency in the grid. An optical power meter measures the transmitted power at the common through port while a multi-objective minimization routine using the NGSAII sampler from the open-source library Optuna \cite{akiba} drives the microheaters appropriately -- even in the presence of mutual thermal crosstalk -- to minimize the transmitted optical power at each frequency position. In this way, individual racetracks are optimally tuned to a common resonance mode and shifted to align with the frequency grid at drive states that give the minimum objective function value. 
For each state of the WSS, filters are first manually tuned to be relatively close to their location on the frequency grid in order to minimize the runtime of the algorithm. A single WSS state is then gathered in roughly 30 minutes using our serially updated electronics. Using an SMU with a faster or parallel update rate would decrease this runtime.

\begin{figure}[t]
\centering
\includegraphics[width=9cm]{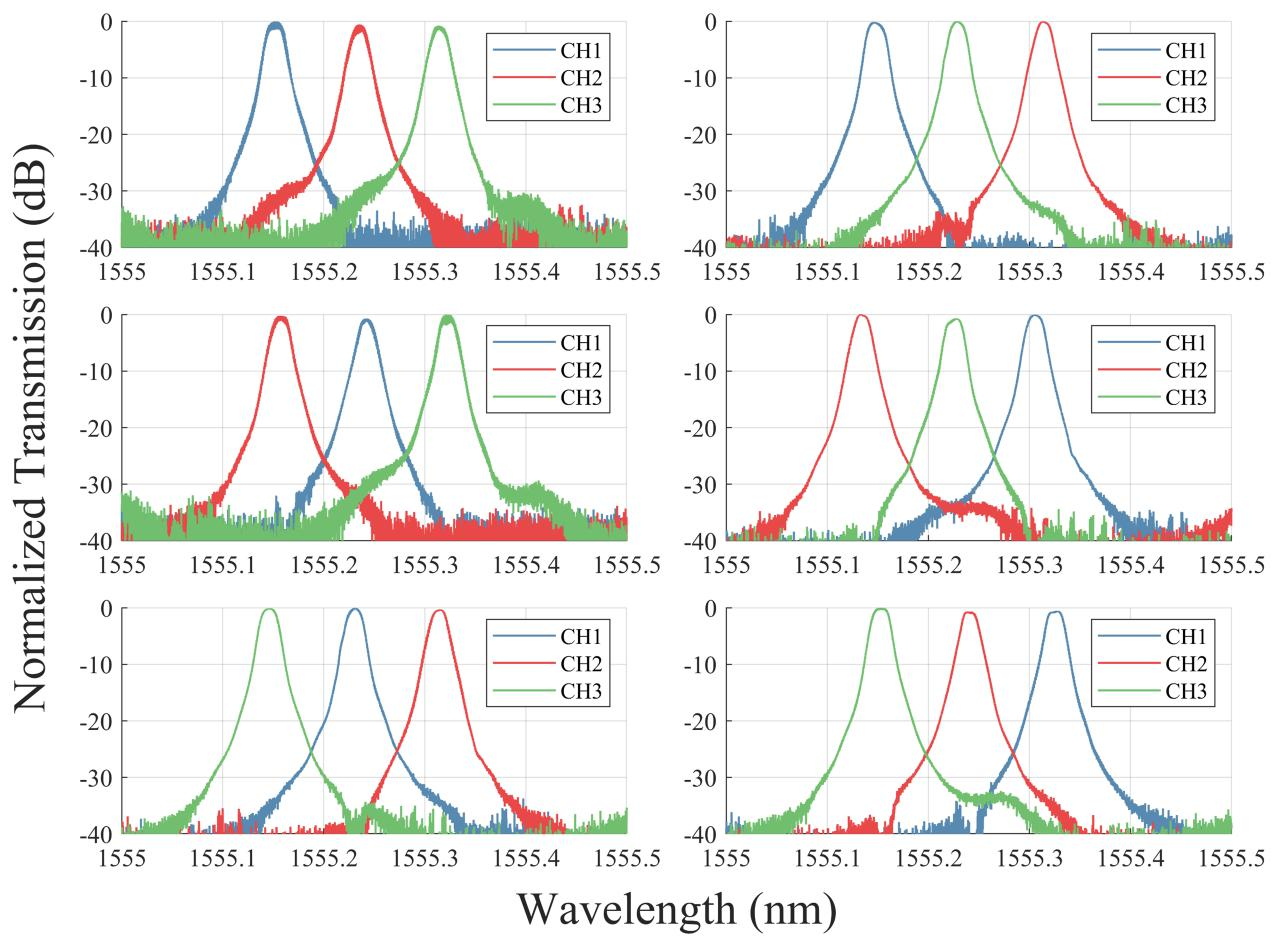}
\caption{ 
Spectra from the six permutations of the three channel WSS as measured from a swept tunable laser source. For each permutation, traces are normalized to the channel with maximum transmission.
}
\label{fig5}
\end{figure}

We programmed a three channel subset of our WSS for operation on a 10 GHz spaced grid near 1550 nm. Each channel was aligned to a unique position on this three-point grid \cite{Cleo_wss}. There are six permutations for three channels on this grid, and the spectra for each state taken from a tunable laser sweep are shown in Fig.~\ref{fig5}. 
From the figure, we can clearly see at the center of any channel better than 25 dB dB of isolation from the (both) adjacent channel(s). The average electrical power consumption for each racetrack is $\approx$~50~mW for thermal tuning. However, racetracks are initially detuned from unused channels to prevent light leakage, contributing overhead to this power budget.

\begin{figure}[t]
\centering
\includegraphics[width=3.5in]{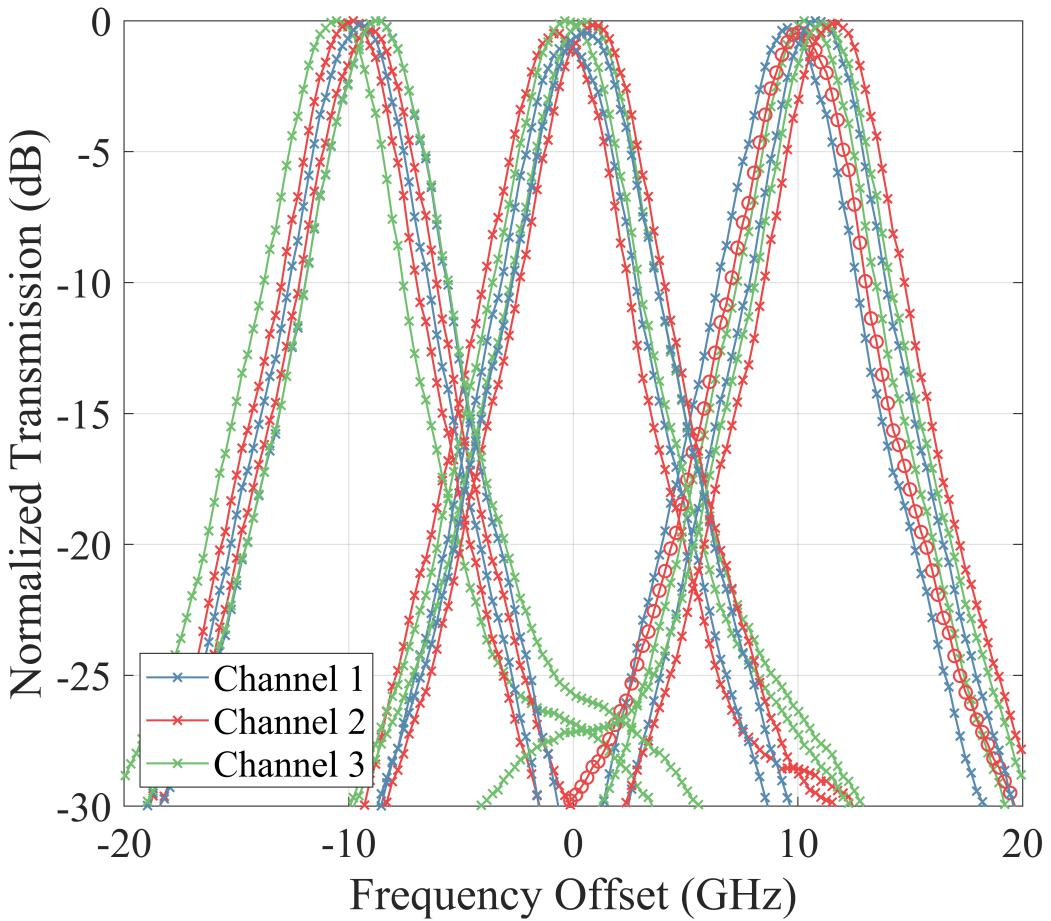}
\caption{ 
Overlayed spectra from the six permutations of the three-channel WSS as measured from an OSA. For each permutation, traces are normalized to the channel with maximum transmission. The input of the WSS is a broadband, flat ASE spectrum from an EDFA.
}
\label{fig6}
\end{figure}

Using the spectra of Fig. \ref{fig5} measured from a scanned laser, we computed an average detuning from the predefined frequency grid of $\approx$~0.8$\pm$0.5~GHz for each channel in the WSS. To avoid possible wavelength registration errors, which can arise during successive wavelength sweeps from a tunable laser, we perform a single-shot measurement over all frequency channels in a permutation with an optical spectrum analyzer (OSA). 
For this purpose, we send a broadband amplified spontaneous emission (ASE) spectrum from an erbium-doped fiber amplifier (EDFA) to the input port of our WSS and measure the output of each channel for all permutations with our OSA. 
An overlay of the measured spectra are shown in Fig. \ref{fig6} \cite{Cleo_wss}. We see a slight increase in optical crosstalk from the OSA measurement that can be attributed to the 0.01~nm ($\approx$~1.247~GHz) resolution of our OSA. 
We compute an average frequency detuning of 0.7$\pm$0.1~GHz for each filter of the WSS using our OSA.
We also measured a loss variation of $<$~0.5~dB from the nominal 1~dB drop-port loss for each channel of the WSS using both laser and OSA spectra.

\begin{figure}
\centering
\includegraphics[width=3.5in]{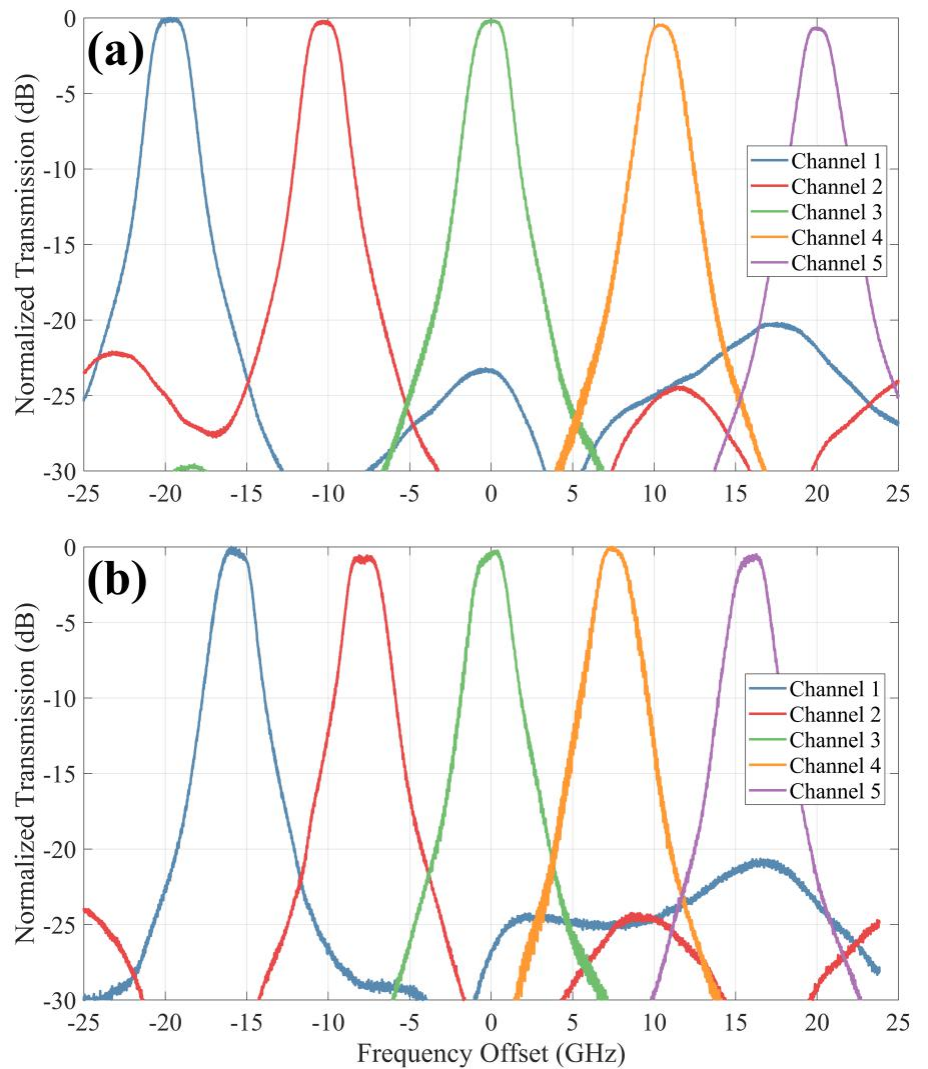}
\caption{ 
Measured spectra for the WSS configured as a 1$\times$5 demultiplexer as taken from a swept laser source for (a) 10 and (b) 8~GHz grid spacing. The center wavelength for both grids is $\approx$~1551 nm. Traces are normalized to the channel with maximum transmission.
}
\label{fig7}
\end{figure}

As a final extension of our device, we program the WSS for a 1$\times$5 demultiplexing operation on both an 8 and 10~GHz frequency grid. Operating the WSS with 5 channels would require 5! optimized permutations, which is possible given the need and enough time. However, here we show only a single permutation as an example. The spectra for both 8 and 10~GHz grids are shown in Fig. \ref{fig7} as measured from a swept tunable laser source. 
Channels 3, 4, and 5 are the channels used in Fig. \ref{fig5} and Fig. \ref{fig6} for WSS operation. We see from the blue and red traces that adding channels 1 and 2 for the demultiplexer results in a slight increase in optical crosstalk at both the high and low-frequency region of the spectrum. Nonetheless, the crosstalk for both 8 and 10~GHz spacing remains below $<$~$-$20~dB in all cases. Besides the crosstalk, we see a transmission variation of $<$~0.75~dB between each channel for both grids.

\section{Discussion}
\label{sec4}

The primary challenges with scaling the high quality devices shown here are: (1) extending the FSR, (2) reducing the electrical power needed for thermal tuning, and (3) developing a robust control solution. Since the quality factor of the racetrack increases with the multimode waveguide length, it is common for the FSR of the high quality devices to be on the order of 10s of GHz, if not lower. A common way of increasing the FSR is by using coupled-resonator structures with different round-trip lengths in a Vernier configuration \cite{schwelb}. This has been shown to extend the operating FSR of the device. Yet it requires significant engineering to get large suppression of the interstitial resonance peaks, among other challenges \cite{boeck}. Likewise adding frequency dependence to the coupling coefficient in the form of an interferometer could function to extend the FSR of the device \cite{watts}. However, an interferometer would add an extra phase tuning element because of fabrication imperfections and thus complicate the operation. 

Reducing electrical power consumption from thermal tuning can be accomplished in a number of ways. For a demultiplexer, frequency positions of each racetrack can be prebiased in fabrication by adjusting the round-trip path lengths. While thermal tuning would still be required to compensate for variation in fabrication, it would take less power than if racetracks were identical like our design here. Still this method is not effective for a WSS since filters need to be tuned to each position on the frequency grid. Second, by thermal isolation trenching of top and bottom oxide materials as well as removing the silicon substrate, the electrically generated heat can be kept from spreading away from the optical waveguide \cite{liu}. Such processes would also help alleviate thermal crosstalk effects further reducing the required electrical power consumption.

A number of control methods could help realize a large-scale on-chip WSS. A common approach to lock microrings to an optical carrier is to monitor the intra-cavity optical power using a photoconductive element embedded within the optical cavity, often in the form of a doped (n, p, or p-n) region of the waveguide \cite{jayatilleka}. However, this fundamentally leads to excess loss thereby reducing the resonator quality factor. Another scheme 
uses a contactless probe placed adjacent to a waveguide capable of sensing light in the waveguide by measuring a conductance change induced by free-carriers being generated at the silicon-insulator interface.
A low-frequency modulation of an optical carrier can then be detected by the probe and used to lock microrings to a laser even in the presence of multiple laser sources \cite{morichetti, aguiar}. In a similar vein, low-frequency thermal modulation of a microring has been used to encode a shallow amplitude modulation onto an optical carrier which can, upon photodetection, induce an asymmetric error signal appropriate for frequency locking \cite{padmaraju}. As monolithic electronic-photonic systems \cite{sun} become increasing available, on-chip control of silicon photonic concepts such as ours may become feasible.

\section{Conclusion}
\label{sec5}

In conclusion, the design methodology and results were presented for a WSS implementing high quality factor microresonator filtering elements. The procedure outlined here can be used on any integrated platform to realize first-time-right devices. By tuning parameters of the subcomponents of the high quality racetrack, arbitrary filter responses can be achieved. For our designed WSS, we demonstrate second-order filters with $\approx$~1~dB of drop-port loss, $\approx$~2~GHz FWHM linewidth, and quick off-resonance roll-off of $\approx$~5.3~dB$/$GHz. We use our system to experimentally show 3 channel WSS operation on a 10~GHz grid and 5 channel multi/demultiplexer operation on both an 8 and 10~GHz grid, both with $<$~$-$~20~dB of inter-channel optical crosstalk. 
A flexible grid spacing and channel count can be accommodated for by applying the appropriate amount of electrical power to each racetrack.
The performance of our devices shows promise towards realizing fine-resolution filtering in future silicon photonic systems.

\section*{Acknowledgments}
This work was funded under NSF grant 2034019-ECCS and by AFRL grant FA8750-20-P-1705 under an STTR through Freedom Photonics.
The authors would like to thank Matthew van Niekerk and Stefan Preble at Rochester Institute of Technology for assistance with wirebonding and package assembly, and Cale Gentry from SRI International for discussions.



\end{document}